\documentstyle[psfig,prd,aps]{revtex}
\def \pom {{I\!\!P}}

\begin{document}
\draft
\title{The Logarithmic Slope in Diffractive DIS}
 \author{M.B. Gay Ducati
$^{1,\dag}$\footnotetext{$^{\dag}$ E-mail:gay@if.ufrgs.br}, V.P. Gon\c{c}alves
$^{2,*}$\footnotetext{$^{*}$ E-mail:barros@ufpel.tche.br}, M.V.T. Machado
$^{1,\star}$\footnotetext{$^{\star}$ E-mail:magnus@if.ufrgs.br}  }
\address{$^1$ Instituto de F\'{\i}sica, Universidade Federal do Rio Grande do
Sul\\
Caixa Postal 15051, CEP 91501-970, Porto Alegre, RS, BRAZIL}
\address{$^2$ Instituto de F\'{\i}sica e Matem\'atica, Universidade
Federal de Pelotas\\
Caixa Postal 354, CEP 96010-090, Pelotas, RS, BRAZIL}
\maketitle

\begin{abstract}
The logarithmic slope of diffractive structure function is a potential
observable to separate the hard and soft contributions in diffraction,
allowing to disentangle the QCD dynamics at small $x$ region. In this paper we
extend our previous analyzes and calculate the diffractive logarithmic slope
for three current approaches in the literature: (i) the Bartels-Wusthoff
model, based on perturbative QCD, (ii) the CKMT model, based on Regge theory
and (iii) the Golec Biernat - Wusthoff model which assumes that the saturation
phenomena is present in the HERA kinematic region. We analyze the transition
region of small to large momentum transfer and verify that  future
experimental results on the diffractive logarithmic slope could discriminate
between these approaches.
\end{abstract}


\bigskip







\section{INTRODUCTION}

Recent measurements of the structure functions for the deep
inelastic $ep$ scattering at HERA has probed the interface between
hard and soft physics in the description of strong interactions.
While the  perturbative QCD DGLAP evolution \cite{DGLAP} describes
very well the experimental results at medium $x$ and high $Q^2$
(respectively, the momentum fraction carried by the partons  and
the momentum transfer), the description of data at small values of
$x$ is a particularly rich and complex subject in which the usual
approach meets and competes with distinct approaches, some based
on QCD, others on Regge theory or essentially {\it ad hoc}
phenomenological models (For a review see e.g. Ref.
\cite{cooper}). Basically these approaches describe very well the
proton structure function data, which implies a theoretical and
experimental challenge: the derivation of a quantity which allows
us to disentangle the hard and soft contributions for the dynamics
at small $x$. Recently, we have proposed the analyzes of the slope
of diffractive structure function as a potential observable to
explicit the leading dynamics at $ep$ diffractive processes
\cite{prl}. Here we study in detail the predictions for this
quantity for three current approaches present in the literature,
which describe very well the  HERA data. Before to present our
results, we discuss the motivation of our work.

About two years ago, the measurement at HERA $ep$ collider of the
logarithmic slope ($Q^2$-slope) of the inclusive structure function $%
F_2(x,Q^2)$ has presented a new challenge for the small $x$ regime \cite
{caldwell,preldata}. Basically, the turn over of the $Q^2$-slope
was initially considered as an evident signal for the change of the dynamics,
namely a transition from the perturbative QCD to the high density regime
(e.g., see Ref. \cite{muelec}). However, adjustments in the usual parton
distribution functions (pdf's) allow to extend the DGLAP formalism to be
valid in lower momentum transfer values \cite{GRV99,MRST99} ($Q^2\leq 3$ GeV$%
^2$ and Bjorken variable $x<5\,.10^{-3}$), with a reasonable
description of the $F_2$ slope data. Furthermore, the HERA results
have motivated the proposition of other approaches which consider the
interplay between hard and soft physics. Essentially, we can
separate these models in two categories: (a) the ones that
assume the Regge framework as starting point at small $x $ and
$Q^2$, and consider improvements to describe the large $Q^2$
region and, (b) those that assume the validity of pQCD at large
$Q^2$ and estimate the perturbative corrections present at small
$x$ and $Q^2$. Although at this moment we have distinct models,
based on different assumptions, which reasonably describe the HERA
data for $F_2$ and its slope, there is the expectation that future
experimental results for the $F_2 $ slope would allow to
discriminate the correct dynamics at small $x$ \cite {levinrec}.

In a similar way as the $F_2(x,Q^2)$ case, the logarithmic slope of the
diffractive structure function should be studied and its role to
disentangle the dynamics (soft and/or hard) needs investigation, since at
present there exists enough statistics in the $F_2^{D(3)}(x_{%
\pom}, \beta, Q^2)$, the diffractive structure function,  experimental
measurements to obtain data for this quantity. In the kinematic domain of the
present experimental measurements, $x_{\pom}$ may be interpreted as the
fraction of the four-momentum of the proton carried by the diffractive
exchange, the Pomeron, if such a picture is invoked. Moreover, $\beta$ is
the fraction of the four-momentum of the diffractive exchange
carried by the parton interacting with the virtual boson.
Theoretically, when we look into diffractive dissociation in deep
inelastic scattering (DDIS), mainly on $F_2^D$, the interplay between hard and
soft regimes is more explicit \cite{Mueller,DDIS}. Basically, the partonic
fluctuations of the virtual photon can lead to configurations of
different sizes when analyzed in the proton rest frame. The study
of the diffractive dissociation of protons has shown that for real
photons ($Q^2 \approx 0$), where the transverse size of the
incoming pair is approximately that of a hadron, the energy
dependence is compatible with the expectations based on the  soft
Pomeron exchange. On the other hand, at large $Q^2$ the energy
dependence is higher than that of the soft Pomeron, suggesting
that pQCD effects may become visible for small incoming
quark-antiquark pairs. Therefore, as the $F_2^D$ structure
function is inclusive to the hard and soft contributions to the
dynamics, the analyzes of its logarithmic slope is
maybe a better test for the Pomeron physics.

First, we consider a Regge inspired model \cite{CKMT1,CKMT2},
where the diffractive production is dominated by the soft
(non-perturbative) Pomeron. In this framework, the correspondent
diffractive structure function is connected with the inclusive one
and a particular expression to the Pomeron flux has been used. In
a second step, one takes into account a pQCD approach \cite{BW},
where the diffractive process is modelled as the scattering of the
photon Fock states ($q\bar{q}$ and $q\bar{q}+$gluon
configurations) with the proton through the color singlet two hard
gluon exchange (in the proton rest frame). Moreover, we calculate
the $F_2^D$-logarithmic slope considering the phenomenological
saturation model from Ref. \cite{GW} and perform a comparison with
the previous non-saturated approach \cite{BW}, pointing out their
main differences. We found that these different approaches predict
very distinct behaviors for the logarithmic slope, which may shed
light into the leading dynamics at $ep$ diffractive processes if
this quantity is experimentally analyzed.

The paper is organized as follows. In the next section we present the basic
formulae and a very brief review of two distinct frameworks for the diffractive
structure function: the Capella et al. (Regge framework) approach without $%
Q^2$ evolution, and the Bartels-W\"usthoff model considering its
phenomenological parameterization (pQCD-based). Moreover, in this
section we derive the expressions for the $F_2^D$-logarithmic
slope. In the Sec. \ref {nosaturation} the correspondent
predictions to the $F_2^D$ logarithmic slope from these models are
presented and discussed. The saturation phenomena is addressed in
Sec. \ref{satu}, where we show the results for both saturated and
non-saturated pQCD approaches. In the last section we draw our
conclusions.
\section{DIFFRACTIVE DIS AND PHENOMENOLOGICAL APPROACHES}

\label{formulae}

\subsection{The Regge inspired model}

Considering diffractive (Ingelman-Schlein) factorization
\cite{Ingelman}, the probe of the virtual photon in DIS provides
understanding on the nature of the Pomeron and the partonic structure of
that object. A few years ago Capella-Kaidalov-Merino-Tran Thanh Van (CKMT)
proposed a model to diffractive DIS based on Regge theory \cite{CKMT1,CKMT2}
and the Ingelman-Schlein ansatz. In this case, deep inelastic diffractive
scattering proceeds in two steps (the Regge factorization): first a Pomeron
is emitted from the proton and then the virtual photon is absorbed by a
constituent of the Pomeron, in the same way as the partonic structure of the
hadrons.

 The Pomeron is considered as a Regge pole with a
trajectory $\alpha_{\pom}(t)=\alpha_{\pom}(0) + \alpha^{\prime}\, t $
determined from soft processes, in which absorptive corrections (Regge cuts)
are taken into account. Explicitly, $\alpha_{\pom}=1.13$ and $%
\alpha^{\prime}_{\pom}=0.25\;GeV^{-2}$. The diffractive contribution to DIS
is written in the factorized form:
\begin{eqnarray}  \label{flux}
F_2^D(x,Q^2,x_{\pom},t)=\frac{[g^{\pom}_{pp}(t)]^2}{16 \pi}x_{\pom%
}^{1-2\alpha_{\pom}(t)}F_{\pom}(\beta,Q^2,t)\;,
\end{eqnarray}
where the details about the coefficients appearing in the Pomeron
flux can be found in Ref. \cite{CKMT2}. 

Moreover, $F_{\pom}$ is
determined using Regge factorization and the values of the triple
Regge couplings obtained from soft diffraction data. Namely, the
Pomeron structure function is extracted
from $F_2^p$, or more precisely from the combination $F_2^d=\frac{1}{2}%
(F_2^p + F_2^n)$, by replacing the Reggeon-proton couplings by the
corresponding triple reggeon couplings (see Ref. \cite{CKMT1}).
One of the main points of this model is the dependence on $Q^2$ of
the Pomeron intercept [$F_{\pom} \propto \beta^{\Delta (Q^2)}$].
For low values of virtuality (large cuts), $\Delta$ is close to
the effective value
found from analyzes of the hadronic total cross sections ($\Delta \sim 0.08$%
), while for high values of $Q^2$ (small cuts), $\Delta$ takes the
bare Pomeron value, $\Delta \sim 0.2-0.25$.

The comparison of the CKMT model with data is quite satisfactory,
mainly when considering a perturbative evolution of the Pomeron
structure function. We notice that here one uses the pure CKMT
model \cite{CKMT1} rather than including QCD evolution of the
initial conditions \cite{CKMT2}, which has been considered to
improve the model at higher $Q^2$. Such procedure ensures that we
take just a  strict Regge model, namely avoiding contamination by pQCD
phenomenology.

Justifying our choice, the CKMT model is a long standing approach
describing in a consistent way  both inclusive and diffractive
deep inelastic based on Regge theory, which is continuously
improved  considering updated experimental results
\cite{Kaidepjc}. In general, Regge inspired approaches focus only
the inclusive case, or exclusively diffractive DIS. A subtle
question concerning the CKMT approach is the dependence of the Pomeron
intercept on the virtuality in the inclusive case.
Although the need of this $Q^2$-dependence, claimed by Kaidalov for a
long time (for a review see \cite{Kaidalovrev}), the smooth interpolation
between a soft and a semihard intercept seems to break off the pure reggeonic
feature of the model.  However, in diffractive deep inelastic the energy
dependence (i.e., the Pomeron flux) is driven by a soft Pomeron with a fixed 
$\alpha_{\pom}=1.13$, properly corrected by absorptive effects. Indeed, this
fact is verified at our previous work \cite{prl}, when considering the
effective slope $\partial \ln F_2^D/ \partial \ln (1/x_{\pom})$. We notice that
here one uses the pure CKMT model \cite{CKMT1} rather than
including QCD evolution of the initial conditions \cite{CKMT2},
which has been considered to improve the model at higher $Q^2$.
Such procedure ensures that we take just a  Regge model, namely
avoiding contamination by pQCD phenomenology.

Considering all properties of the diffractive structure function
on the CKMT model, the calculation of the logarithmic slope is
straightforward. The expression reads now:
\begin{eqnarray}
\frac{dF_2^D}{d\,\ln Q^2} =  {\cal N}\,\,
x_{\pom}^{1-2\,\alpha_{\pom}(0)}\, \left[ \, eA \, \, \beta^{-
\Delta(Q^2)}\,(1 - \beta)^{n(Q^2)\,+\,2}\, \left( {\frac{Q^2 }{Q^2
+ a}} \right )^{1\, +\, \Delta(Q^2)}\,S_{\pom}(Q^2,\beta)\,
 \,\, + \,\, \right.  \nonumber \\
 \left.  \,fB \, \beta^{1 - \alpha_R} \, (1 -
\beta)^{n(Q^2)\,-\,2} \,\left( {\frac{Q^2 }{Q^2 + b}}
\right)^{\alpha_R} \, S_R(Q^2,\beta)\, \right]& & \,\,,
\end{eqnarray}
\noindent where the overall normalization ${\cal N}$ comes from
the integration over $t$ of the Pomeron flux, $n(Q^2) = {\frac{3
}{2}} \left ( 1 + {\frac{Q^2 }{Q^2 + c}} \right )$, $\alpha_R$ is
the  secondary reggeon intercept (CKMT supposes that the only
secondary trajectory that contributes is the $f$ one).  The coefficients and
constants are taken from   Ref. \cite{CKMT2}.
Moreover, the factors $S_{{\pom}%
,R}(Q^2,\beta)$ are defined as
\begin{eqnarray}
S_{\pom}(Q^2, \beta) & = & \Delta(Q^2)\,\left[\, \frac{a}{Q^2 + a}\,
\right]\,+\,\frac{3\,c}{2}\,\left[\, \frac{Q^2}{(Q^2+c)^2}%
\,\right]\,\ln(1-\beta) \, +   \nonumber \\
& + & \, 2\,d\,\Delta_0\,\left[\,\frac{Q^2}{(Q^2 + d)^2} \,\right] \, \ln
\left( \frac{Q^2}{\beta\,(Q^2 + a)} \right) \,\,, \label{sp} \\
S_R(Q^2,\beta) & = & \,\, \alpha_R(0) \, \left[ \, \frac{b}{Q^2 + b} \,
\right] \, + \, \frac{3\,c}{2}\,\left[ \, \frac{Q^2}{(Q^2+c)^2}
\,\right]\,\ln(1-\beta) \,\,. \label{sr}
\end{eqnarray}

The above expressions have been used in our previous work \cite{prl} to
obtain the main features of this model concerning the $F_2^D$-logarithmic
slope, analyzing the $x_{\pom}$, $\beta$ and $Q^2$ spectra under a
kinematical constraint which relates the variables $x$ and $Q^2$. In the
next section we will present a brief review of the results from Ref. \cite
{prl} and a full analyzes of this quantity when the kinematic constraint is
not assumed. Before, however, we present the derivation of the $F_2^D$
slope in the pQCD approach.

\subsection{The perturbative QCD approach}

\label{pQCDsec}

Although the diffractive dissociation is mainly connected to soft
processes and thus linked with the Regge theory, the pQCD framework
has been recently used by some authors to describe quite well the diffractive
structure function \cite {MW,authors}. The main properties of the pQCD models 
are very similar. In particular, we consider for our analyzes the
Bartels-Wusthoff model and its further parameterization to the 
measurements \cite{BW}. The physical picture is that, in the
proton rest frame, diffractive DIS is described by the interaction
of the photon Fock states ($q\bar{q}$ and $q\bar{q}g$
configurations) with the proton through a Pomeron exchange,
modeled as a two hard gluon exchange. The corresponding structure
function contains the contribution of $q\bar{q}$ production to
both the longitudinal and the transverse polarization of the
incoming photon and of the production of $q\bar{q}g$ final states
from transverse photons. The basic elements of this approach are
the photon light-cone wave function and the non-integrated gluon
distribution (or dipole cross section in the dipole formalism).
For elementary quark-antiquark final state, the wave functions
depend on the helicities of the photon and of the (anti)quark. For
the $q\bar{q}g$ system one considers a gluon dipole, where the
$q\bar{q}$ pair forms an effective gluon state associated in color
to the emitted gluon and only the transverse photon polarization
is important. The interaction with the proton target is modeled
by two gluon exchange, where they couple in all possible
combinations to the dipole. In a comparison with data, the
transverse $q\bar{q}$, $q\bar{q}g$ production
and the longitudinal $q\bar{q}$ production dominate in distinct regions in $%
\beta$, namely medium, small and large $\beta$ respectively \cite{BW}. The $%
\beta$ spectrum and the $Q^2$-scaling behavior follow from the
evolution of the final state partons, and are derived from the
light-cone wave functions of the incoming photon, decoupling from
the dynamics inside the Pomeron, while the energy dependence and
the normalizations are free parameters.

The calculation of the expression for the $F_2^D$-logarithmic
slope is straightforward, considering each contribution coming
from the different configurations of the photon Fock state. Moreover, this approach allows to obtain
parameter free predictions for the logarithmic slope, since all
parameters have been obtained in comparison with the H1 and ZEUS
data. We justify our choice due to the simple analytic expressions for the
diffractive structure function for each Fock state configuration
which turn out the analyzes clearest and avoiding cumbersome
numeric calculations.

 The explicit expressions for the diffractive logarithmic
slope are written as
\begin{eqnarray} \frac{d\,F_2^{D(3),\,q\bar{q}T}(x_{\pom}, \beta, Q^2)}{d\,\ln
Q^2} &=& \frac{n^1_2}{ (\, \ln \frac{Q^2}{Q^2_0} +
1\,)}\,\,F_2^{D(3),\,q\bar{q}T}(x_{\pom}, \beta, Q^2) \,\,,
\label{dpqcd} \\
\frac{d\,F_2^{D(3),\,q\bar{q}G}(x_{\pom}, \beta, Q^2)}{d\,\ln Q^2}
&=& \frac{1}{(\, \ln \frac{Q^2}{Q^2_0} + 1\,)}\,\left[\, n^1_2
\,+\, \frac{Q^2}{Q^2+Q^2_0}
\,\right]\,\,F_2^{D(3),\,q\bar{q}G}(x_{\pom}, \beta, Q^2) \,\,,
 \nonumber \\
\frac{d\,F_2^{D(3),\,q\bar{q}L}(x_{\pom}, \beta, Q^2)}{d\,\ln Q^2}
&=& \left[ \frac{n^1_4}{(\,\ln \frac{Q^2}{Q^2_0} + 1\,)} \,+\,
\frac{Q^2 - 7\,\beta Q^2_0}{Q^2 + 7\,\beta Q^2_0}
\,\right]\,\,F_2^{D(3),\,q\bar{q}L}(x_{\pom}, \beta, Q^2) \,\,.
\nonumber
\end{eqnarray}
 The expressions of the distinct
contributions to $F_2^{D(3)}$ can be found in Ref. \cite{BW}, as
well as the remaining parameters.

The above expressions describe the
$x_{\pom}$, $\beta$ and $Q^2$ behavior of the $F_2^D$-logarithmic slope in the
pQCD model. In next section we present a brief review of the main conclusions
from Ref. \cite{prl} and extend our analyzes for the case where a kinematical
constraint is not assumed.

\section{THE F$_2^D$ LOGARITHMIC SLOPE FROM REGGE AND PQCD-BASED APPROACHES}

\label{nosaturation}

Before we perform the analyzes of the presented models, some comments are in
order. In the first extraction of the $F_2$ slope data at HERA, the
so-called Caldwell plot \cite{caldwell}, the variables $x$ and $Q^2$ were
strongly correlated. For a limited acceptance (in the case of the latest
HERA $F_2$-slope data) and for a fixed energy, one always has a limited range
in $Q^2$ at any given $x$, with average $<Q^2>$ becoming smaller for smaller 
$x$. The latest measurements of the $F_2$-slope presents an increase when $
x$ decreases down to $\sim 10^{-4}$, and then a turn over appears, with a
slope decreasing at lower $x$ (or $Q^2$) values. This peak is currently
interpreted as a transition region from a  a perturbative
QCD regime to a Regge behavior and the maximum is related to the
correlation between the kinematical variables mentioned above.

$ $ From the theoretical point of view however, the logarithmic slope depends
on two variables ($x$ and $Q^2$) which are independent and one is not
restricted to follow a particular path on the surface representing the $Q^2$%
-slope. However, to perform a comparison with those data, due to low
statistics, we should connect $x$ and $Q^2$ by some analytical dependence
$Q^2=Q^2(x)$ that lies within a physical region on ($x ,\,Q^2$)-plane. This
region is bounded by the condition 
\begin{eqnarray}
y=\frac{Q^2}{x\,(s-m^2_p)}\leq 1 \,\,.
\end{eqnarray}
For the HERA experiments, the centre of mass energy is $\sqrt{s}\approx 300$
GeV and this condition writes $Q^2 < 9.10^4\,x$. An usual procedure is to
perform a fit to the ($x,\,Q^2$)-plane satisfying the above condition. In
the previous work \cite{prl}, one considers the path used by the MRST99
group \cite{MRST99} in their analyzes of the $F_2$-logarithmic slope, $%
Q^2=1280\,x^{0.705}$ (for H1 slope data). We take this into account because,
concerning the $F_2^D$-slope, in a first trial study probably we should expect 
a poor statistics in a similar way as it was for the inclusive case \cite
{caldwell}.

\begin{figure}[t]
\centerline{\psfig{file=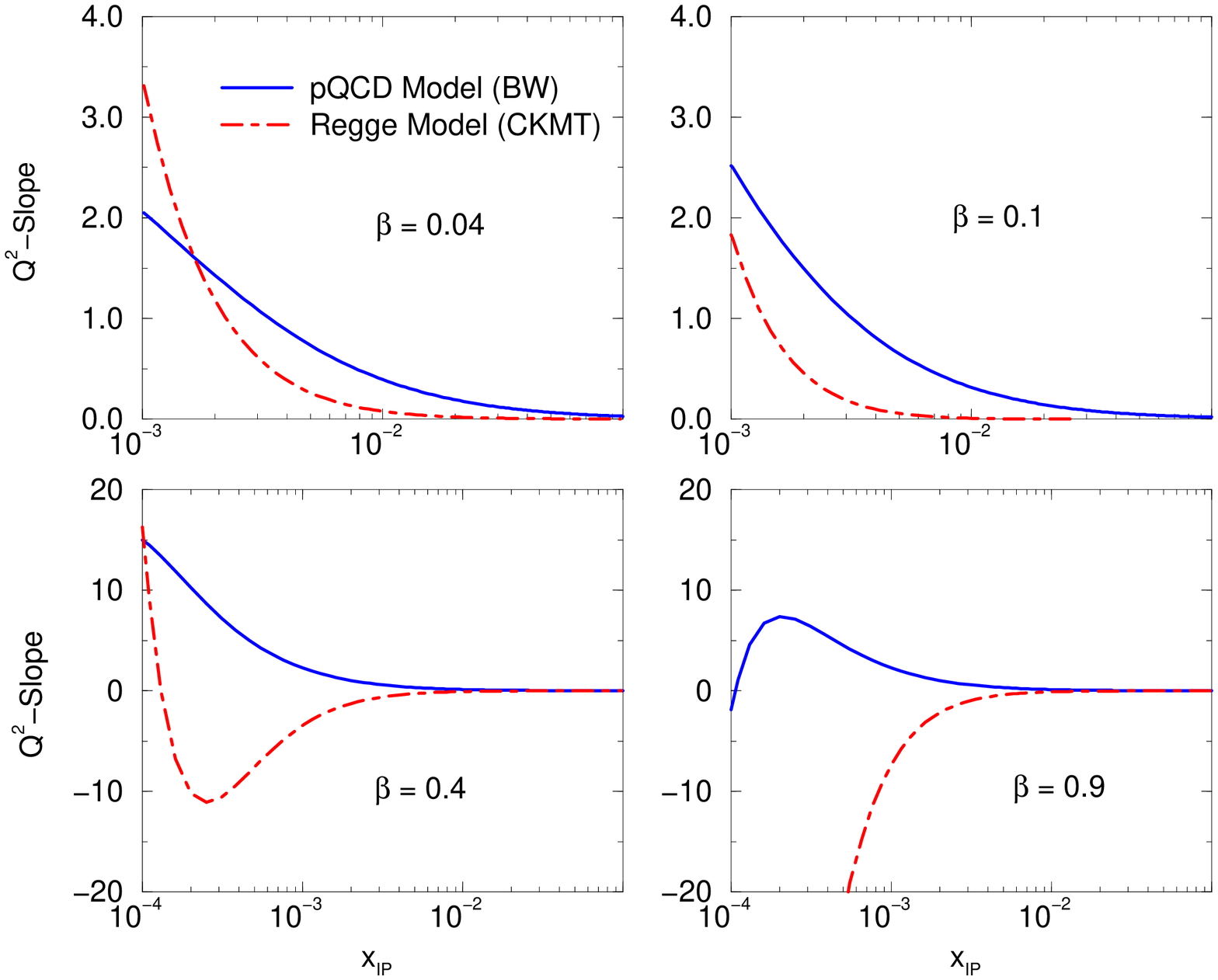,width=150mm}} 
\caption{The $Q^2$-slope versus
$x_{\pom}$ for the pQCD approach (solid lines) and Regge-CKMT 
(dot-dashed lines), at $\beta$ typical values. We consider a kinematical
constraint, $Q^2=Q^2(x)$, taken from MRTS99.}
\label{fig4}
\end{figure}

In Fig. \ref{fig4} one shows the quantity $dF_2^{D(3)}/d\,\ln (Q^2)$
(shortly, $Q^2$-slope) versus $x_{\pom}$ at $\beta $ typical values,
considering the kinematical constraint discussed above. Due to the
kinematical constraint we show the results at $\beta $=0.04, 0.1, 0.4 and
0.9 starting at $x_{\pom}=10^{-3}$, meaning virtualities $Q^2>1$ GeV$^2$ (in
which the pQCD model is valid). We also do not consider the introduction of
reggeonic contributions, which are important at low $\beta $ (larger $x_{\pom%
}$).

We found that the CKMT model (without 
$Q^2$ evolution) provides a transition between positive and negative slope
values at $\beta=0.4$, when considering the $x_{\pom}$ dependence. This
behavior is consistent since the Pomeron structure function in this model is
related to the nucleon structure function $F_2$, which presents that feature
due to the scaling violation. Moreover, in Ref. \cite{prl} we have verified
that the $\beta$ dependence, at $x_{\pom}\ge 10^{-3}$, is predominantly flat
on the whole kinematic range of $\beta$ and it has a turn over in $\beta=0.1$
at $x_{\pom}=10^{-4}$. Since the Pomeron structure function in the CKMT is
just the inclusive structure function $F_2$, the large value of the slope at
small $\beta$ (corresponding to the Bjorken $x$ in inclusive DIS) was
expected, as well as the presence of a similar turnover found in the first
measurements of the inclusive structure function slope. Considering the DGLAP
evolution \cite{CKMT2}, we would expect a $Q^2$-slope close to zero at $%
\beta=0.65$ and even negative at $\beta=0.9$. An important point is the
faster decreasing of CKMT at large $\beta$, coming from the logarithmic
factors $S_{{\pom},\,R}(Q^2,\beta)$. Furthermore, the CKMT model predicts a
constant $x_{\pom}$-slope. This fact comes from the choice of the Pomeron
flux, which does not present an effective intercept depending on virtuality $%
Q^2$, contrasting with the inclusive case where the intercept smoothly
interpolates the soft and semi-hard regions. Further, we return to the
analyzes of the CKMT model, disregarding the kinematical constraint, in
explicit comparison with the pQCD approach reviewed in the subsection (\ref
{pQCDsec}).

Concerning the pQCD model, we have that the slope is predominantly positive
in almost all  considered
$\beta$, taking negative values only at $\beta = 0.9$ for the
interval $x_{\pom}<0.0004$. As shown in Ref. \cite{prl}, the slope presents
a $\beta$ dependent turn over, which is shifted to greater $x_{\pom}$ values
as $\beta$ increases. The positive behavior of the slope at low values of $%
\beta$ is associated to the $q\bar{q}G$ contribution, while for intermediate %
$\beta$ the $q\bar{q}T$ state dominate, producing an almost constant
function on $Q^2$. The high $\beta$ behavior is consistent with the H1
measurements, in the region $x_{\pom}>10^{-3}$, which prefer a positive
slope in $Q^2$, corresponding to a large $q\bar{q}G$ contribution in this
region \cite{BW}. Furthermore, we have shown that the $\beta$ dependence, at %
$x_{\pom}\ge 10^{-3}$, is predominantly flat on the whole kinematic range of %
$\beta$ and it has a turn over at $\beta=0.5$ and $x_{\pom}=10^{-4}$.
Moreover, this model gives a $x_{\pom}$-slope dependent on both the
virtuality $Q^2$ and $\beta$. The virtuality dependence is clear from the
parametrization on the  energy, constructed to interpolate between a softer
effective intercept at low $Q^2$ up to semihard values at higher
virtualities. The hard-like value of the intercept at higher $\beta$ is a
feature already discussed in Refs. \cite{BW}.

Confronting the approaches, we conclude in \cite{prl} that both models
predict a positive slope up to $\beta \sim 0.4$, with a steeper decreasing
in CKMT. The high $\beta $ region discriminates the behaviors. The pQCD
results provide a positive slope, while CKMT produces negative values. This
comes from the fact that  the CKMT approach does not include the $q\bar{q}G$
contribution, which is dominant in this region (small $x_{\pom}$ and $Q^2$) 
for the pQCD model. The $Q^2$-behavior in the CKMT is determined by the $F_2$
scaling violations, including only  the $q\bar{q}_{T,\,L}$
contributions.

\begin{figure}[t]
\centerline{\psfig{file=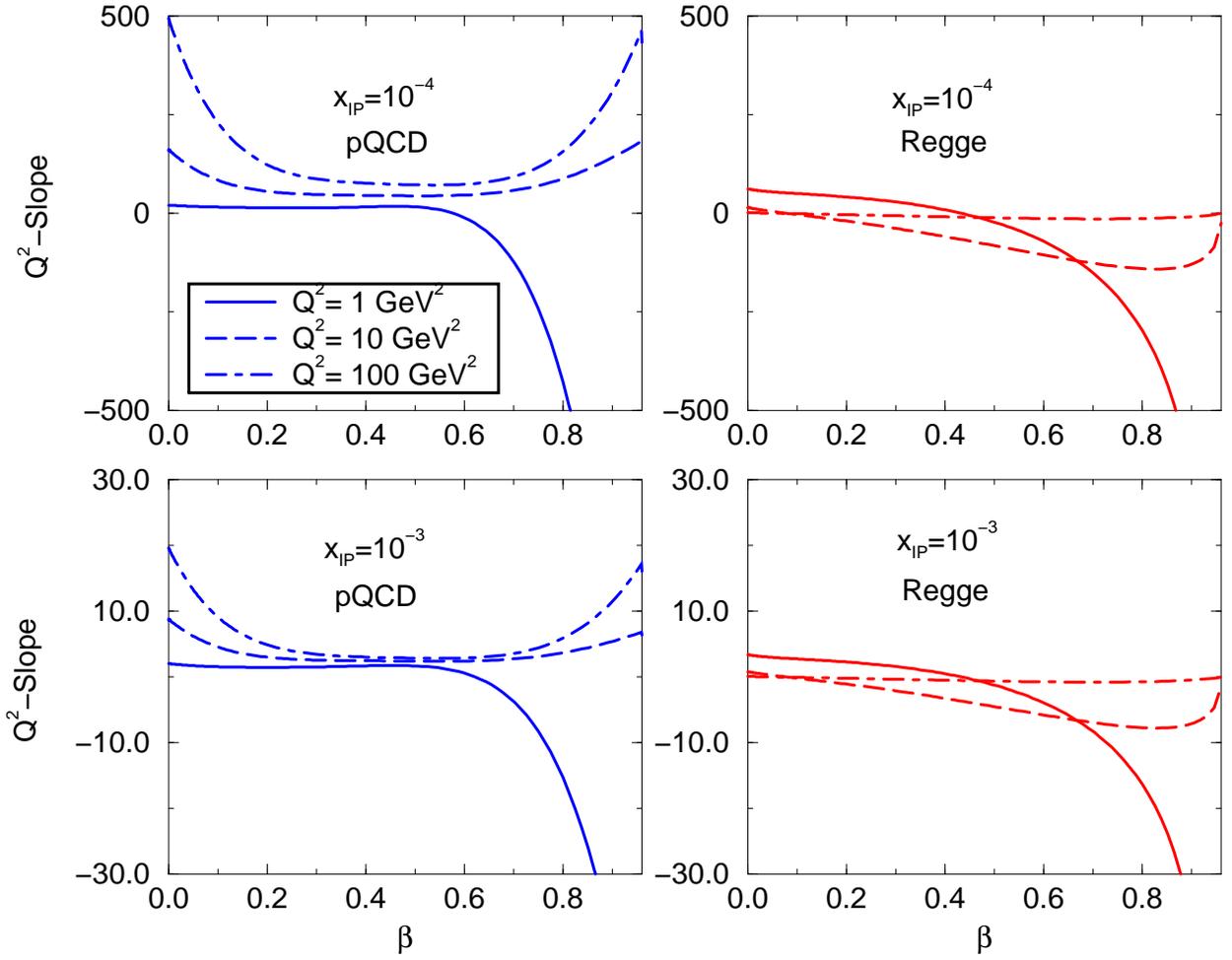,width=150mm}}
\caption{The $Q^2$-slope versus
$\beta$ for the pQCD approach (on the left) and CKMT model (on the right) without
kinematical constraint.  The graphs present the result for fixed $x_{\pom}$ and typical
$Q^2$ values.}
\label{fig5}
\end{figure}

\begin{figure}[t]
\centerline{\psfig{file=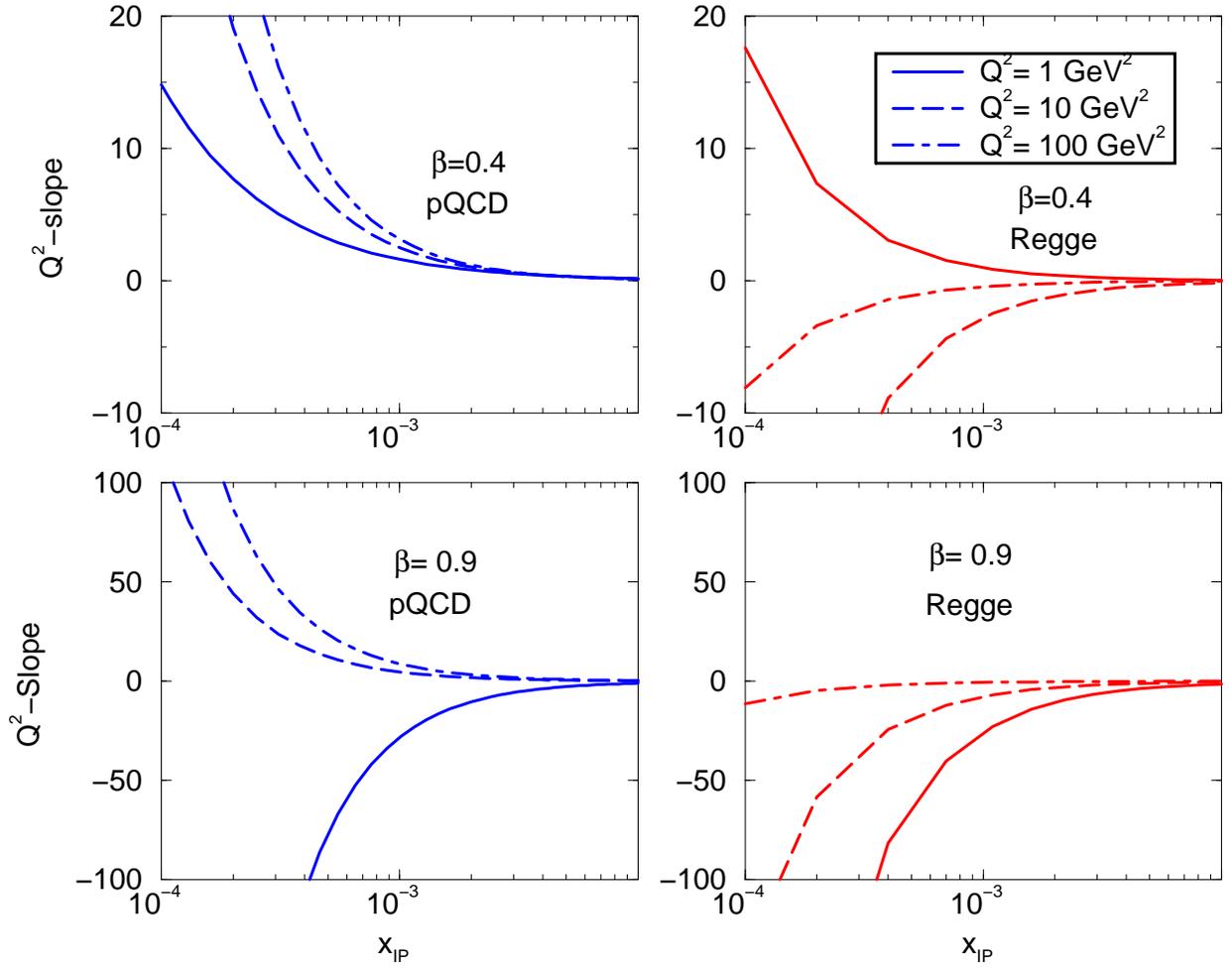,width=150mm}} \caption{The $Q^2$-slope versus
$x_{\pom}$ for the pQCD approach (on the left) and CKMT model (on the right) without
kinematical constraint. The low $\beta$ values were excluded to avoid the dominant
reggeonic contribution in this region.} \label{fig6}
\end{figure}

In Figs. \ref{fig5}-\ref{fig6} we present a new comparison between the
models,  without imposing a kinematical constraint. In Fig. \ref{fig5}
one shows the $\beta $-dependence for typical values of $x_{\pom}$ and $Q^2$%
, where the momentum transfer is ranging from 1 up to 100 GeV$^2$. The CKMT
model predicts a flat behavior on the whole $\beta $ range. Particularly, at %
$Q^2=1\,\,GeV^2$ there is a strong decreasing of the slope at large $\beta $%
. This is due to the presence of factors $\ln (1-\beta )$ in the second
terms of Eqs. (\ref{sp}-\ref{sr}). For larger $Q^2$ values, the logarithm on %
$Q^2$, namely the third term of Eq. (\ref{sp}), compensates the decreasing.
At momentum transfer of 100 GeV$^2$, the CKMT predicts a flat behavior  of
the slope into all  $\beta $ spectrum. The pQCD model produces an increasing
of the slope in both small and large values of $\beta $, while presents a
flat behavior at the medium one. These increasing on the slope are due to
the enhancements in the $q\bar{q}G$ (dominant at low $\beta $) and $q\bar{q}L%
$ (leading at high $\beta $) contributions from the $Q^2$-factors appearing
in Eq. 5 ($q\bar{q}G$ and $q\bar{q}L$), respectively. A steep $Q^2$-slope
decreasing into negative values is also present at the virtuality $%
Q^2=1\,\,GeV^2$ for large $\beta $, in a similar way to the CKMT model.
However, betond the contribution of the (dominant) longitudinallly polarized
$q\bar{q}$ pair configuration, this region also receives contributions
associated to the $q\bar{q}G$ configuration, as discussed above. 

In Fig. \ref{fig6} one presents the $x_{\pom}$ dependence at medium and
large $\beta $. The low $\beta $ region was disregarded in order to avoid to
deal with subleading reggeonic contributions, important in this kinematical
domain. We observe again a smooth behavior for the pQCD predictions, with a
positive slope in almost the whole range (negative values are present at
both large $\beta $ and $Q^2=1\,GeV^2$, in agreement with the discussion in
the previous paragraph). The CKMT model predicts predominantly negative
slope values in this kinematical domain, converging to a flat value for
larger $x_{\pom}$. A clear difference between the predictions of the two
models is the change of signal of the slope with the $Q^2$ evolution at
medium $\beta $ present in the CKMT\ prediction. Those behaviors are also
consistent with the previous plots using the kinematical constraint.

In summary, the results above should allow to discriminate the behaviors
predicted from the different approaches, namely perturbative QCD (hard
physics) and non-perturbative (soft) physics. Theoretically, this difference
comes from the ansatz for the photon-proton interaction (hard or soft) and
also from the relation of the Pomeron structure with the inclusive structure
function used in the CKMT model. This relation implies the inclusion at most
of the 
$q\overline{q}_T$ and $q\overline{q}_L$ configurations in the photon wave
function in its estimates. On the other hand, the pQCD model analyzed here
also  includes the contribution of gluon emission in the photon wave function,
which dominates at small $\beta $. Therefore, the analyzes of the $\beta $
spectrum of the $F_2^D$ slope would be important in an experimental
studies. Moreover, concerning to the $x_{\pom}$ spectrum, the signal of the
slope at medium values of $Q^2$ ($Q^2\approx 10$ GeV$^2$) is a good search
of information about the dynamics. In comparison with the 
experimental measurements of $F_2^D$, the results have favored the
perturbative QCD framework \cite{DDIS}.

\section{THE DIFFRACTIVE LOGARITHMIC SLOPE AND THE SATURATION PHENOMENA}

\label{satu}

In this section we take into account the saturation phenomena, considering
its implications in the analyzes of the
$F_2^D$-slope. The concept of saturation in DIS is connected to the
transition in the underlying physical picture from high to low photon
virtuality $Q^2$, which is present in the measurements of the total $%
\gamma^* p$-cross section at HERA. From the theoretical point of
view, saturation at low momentum fraction $x$ occurs in the
semi-hard region in which the parton density becomes quite high and
the recombination effects tame the growth of the density
\cite{satmodels}. As a consequence, the increasing of the total
cross section is diminished at small $x$.

In perturbative QCD, in the proton rest frame, the $\gamma ^{*}p$
process is described in terms of the photon splitting into a
quark-antiquark pair, far upstream of the nucleon, which then
scatters in the proton \cite{nikolaev} [For a review, see e.g.
Ref. \cite{forshaw}]. In the perturbative regime this reaction is
mediated by the one gluon exchange which turns out into a
multi-gluon one when the saturation region is approached. A
remarkable feature is that the mechanism leading to the photon
dissociation and the further scattering has a factorizable form,
written in terms of a convolution between a photon wave function
and a quark-antiquark cross section \cite{nikolaev}. The wave
function to be calculated from perturbation theory, or the
quark-antiquark cross section, so-called dipole cross section
$\hat{\sigma}$, has a strong influence of the non-perturbative
contributions and should be modeled. A review on issues
concerning the modelling of the dipole cross section can be found
in Ref. \cite{amirin}. Here, we choose the saturation model from
Ref. \cite{GW}, which reproduces the experimental results at both
inclusive and diffractive electroproduction. 

Some comments about
the saturation  model are in order. Although this approach describes in very
good agreement both DIS and diffractive dissociation, a more deep
understanding of the underlying dynamics is  far from clear.
Namely, a connection with the gluon distribution function and the
mechanism of unitarity control is needed (for a review of the
current theoretical effort see, e.g., Ref. \cite{levsat}).
Moreover, the impact parameter dependence ($b_t$) of
the process is not taken into account and the  connection with
DGLAP approach is missing (see criticisms to these points at \cite{LevDiff}).

In that phenomenological approach, the dynamics of saturation is
present in the effective dipole cross section, modelled in an
eikonal way:
\begin{eqnarray}
\hat{\sigma}(x,\,r^2)=\sigma _0\,\left[ \,1-{\rm exp}\left( -\frac{r^2}{%
4\,R_0^2(x)}\right) \,\right] \,\,,
\end{eqnarray}
\noindent
where the $x$-dependent saturation scale (the saturation radius)  is given
by
\begin{eqnarray}
R_0(x)=\frac 1{Q_0}\,\left( \frac x{x_0}\right) ^{\lambda /2}\,\,.
\end{eqnarray}
The normalization $\sigma _0$ and the parameters $x_0$, $\lambda >0$ were
determined from all inclusive data on $F_2(x,Q^2)\sim \sigma _T(x,Q^2)$ with %
$x<10^{-2}$ \cite{GW}. The constant $Q_0$ sets the dimension. In summary,
this phenomenological model consistently interpolates between the saturation
regime and the scaling regime of $\sigma _T\,(F_2)$.

 In
order to calculate the logarithmic slope from the above approach,
we should consider the diffractive structure function
$F_2^{D(3)}$, which is given by \cite{BW}
\begin{eqnarray}
F_2^D(x_{\pom},\beta ,Q^2)\sim \beta \,\int \,dt\,\int \,\frac{k_t^2\,d^2k_t%
}{(1-\beta )^2}\,\,\left| \int \,\frac{d^2l_t}{l_t^2}\,D\Psi (\alpha ,k_t)%
{\cal F}(l_t^2,k_0^2;x_{\pom})\right| ^2\;,
\end{eqnarray}
where $D\Psi $ is a combination of the concerned wave functions,
$l_t$ is
the transverse momentum of the exchanged gluons and ${\cal F}(l_t^2,k_0^2;x_{%
\pom})$ defines the Pomeron amplitude (non-integrated gluon
distribution).
 To take saturation into
account, in Ref. \cite{GW} the unintegrated gluon distribution
${\cal F}(x,l_t^2)$ was calculated from the effective dipole cross
section (the notation is the same of Sec. \ref{pQCDsec} ) as
\begin{eqnarray}  \label{sig_f}
\hat{\sigma}(x,r^2)&=&\frac{4\, \pi^2}{3}\; \int \frac{d l^2_t}{l_t^2} \;
\left[\;1-J_0(l_t r)\;\right]\;\alpha_s\,{\cal F}(x,l_t^2)\;\;,
\end{eqnarray}
with
\begin{eqnarray}
\alpha_s\,{\cal F}(x,l_t^2)&=&\frac{3 \;\sigma_0}{4 \pi^2} \;
R_0^2(x)\;l_t^2\; e^{-R_0^2(x)\, l_t^2} \;\;.
\end{eqnarray}
The corresponding diffractive structure function is constructed
summing three terms coming from the diffractive production of a
quark-antiquark pair (photon polarized transverse and
longitudinally) plus the contribution of the pair with emission of
an additional gluon in the final state. The gluon emission is
known only in certain approximations: transverse (valid at very
large virtualities and finite $M_X$) or longitudinal (at large
$M_X$ and finite virtualities) momenta components strongly
ordered. This  approach described above reproduces accurately the
$\beta$ and $x_{\pom}$ distributions from the H1 and ZEUS data
\cite{cooper}, with a free parameter description \cite{GW}. The main
characteristic is the use of an universal non-integrated gluon
distribution, modelled in a simple way to take into account the
saturation phenomena and completely determined from the inclusive
experimental data.

\begin{figure}[t]
\centerline{\psfig{file=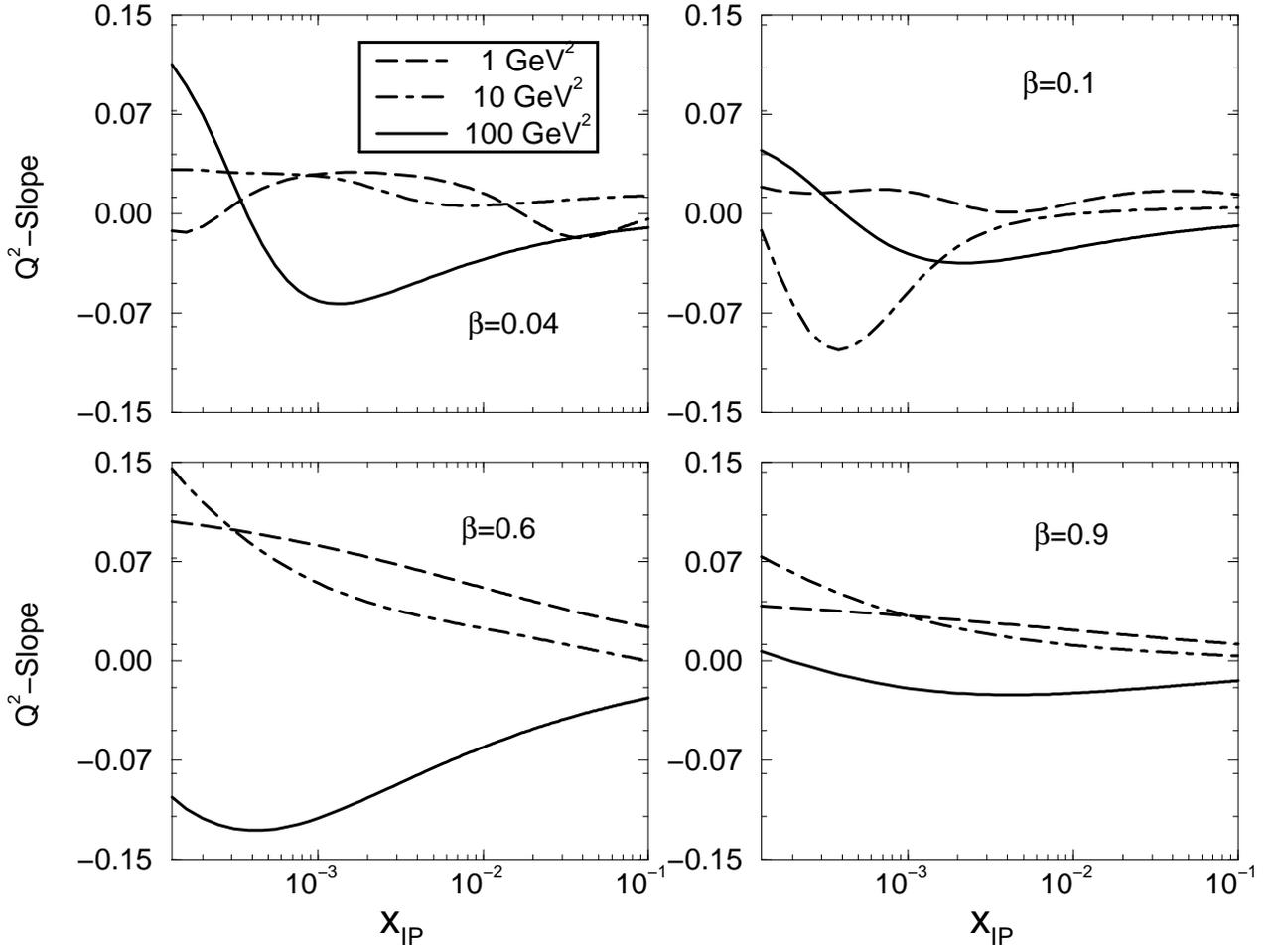,width=150mm}}
\caption{The $x_{\pom}$ dependence of
the logarithmic slope from the saturation model  for the diffractive
structure function, presented at typical $\beta$ values. The  analyzes is
performed on photon momentum transfer $Q^2$ ranging from 1 up to 100 GeV$^2$.}
\label{slxp}
\end{figure}

Now, having reviewed the main formulae concerning the phenomenological
saturation model, we perform numerically the
$F_2^D$ logarithmic slope, using the same parameters from Ref. \cite{GW}. In
this section we denote $Q^2$-slope as the function $x_{\pom}\,dF_2^D/d\ln Q^2%
$. Since the $q\bar{q}G$ configuration dominates at small $\beta$, while the %
$q\bar{q}_{T}$ one dominates at medium $\beta$ and the
$q\bar{q}_{L}$ is important in the large $\beta$ region,  the
behavior of the slope presents a strong dependence in this
variable. In Fig. \ref{slxp}, we show the $x_{\pom}$ dependence,
considering typical $\beta$ values and virtualities ranging from 1
up to 100 GeV$^2$. A remarkable fact is the presence of a
transition between positive and negative slope signals, instead of
a predominantly positive slope as in the non-saturated case. The
transition is not present in the pQCD model without saturation due
to the assumptions made in the parameterization of the HERA data
in the region of large $Q^2$ (large $\beta$). It is important to
emphasize that a transition is
verified in the preliminary ZEUS analyzes of diffractive DIS \cite{prelZEUS}%
, where the saturation model \cite{GW} is considered to describe the $Q^2$
dependence of the diffractive structure function, using as kinematical
variables $M_X$ and $W$ rather than $\beta$ and $x_{\pom}$. Such a procedure
is performed due to the similarity of the behavior of $d\sigma/dM_X$ and $%
\sigma_{tot}(\gamma^*p)$ in the same kinematic range. In that analyzes, the
growth of $x_{\pom}F_2^D$ versus $Q^2$ is stopped at $Q^2\sim 10$ $GeV^2$
and has a smooth decreasing for larger virtuality values. The transition
region corresponds to $\beta \sim 0.2$ for $M_X=5$ GeV and $\beta \sim 0.07$
for $M_X=11$ GeV.

\begin{figure}[t]
\centerline{\psfig{file=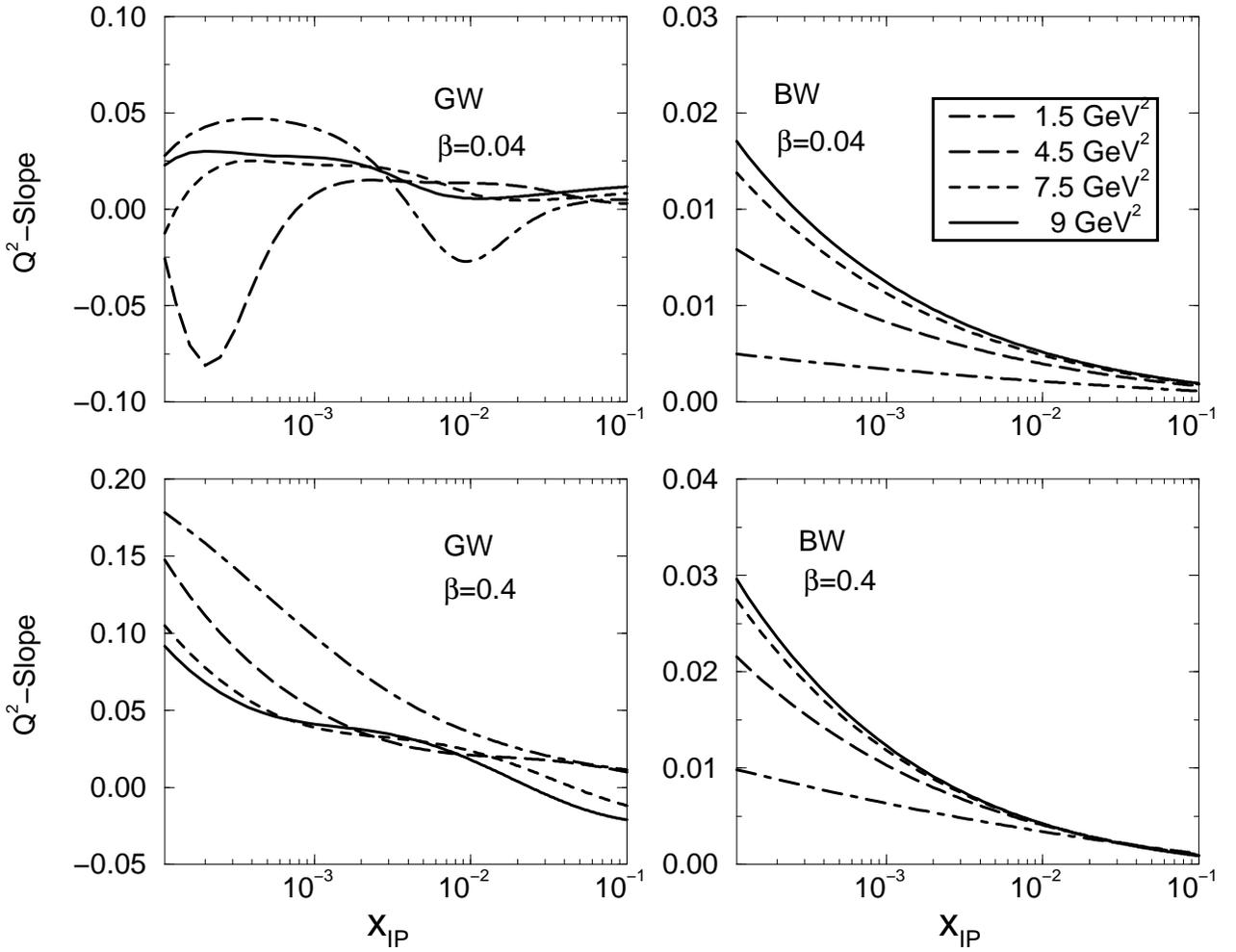,width=150mm}}
\caption{The $x_{\pom}$ dependence
on the logarithmic slope  from both Bartels-Wusthoff model  (denoted BW, on
the right) and the Golec Biernat-Wusthoff Saturation model  (denoted  GW,
on the left), presented at typical $\beta$ values. The  analyzes is performed
on low vitualities $Q^2$ ranging from 1.5 up to 9 GeV$^2$.}
\label{gwcompxp}
\end{figure}

In Fig. \ref{gwcompxp}, one shows the $x_{\pom}$ behavior for both
approaches with \cite{GW} and without \cite{BW} saturation, at typical $\beta%
$ values. We analyze in particular the transition region between hard and
soft dynamics, settled by the low virtualities $Q^2 \sim 1.5 - 9$ GeV$^2$.
The saturation model produces a transition between positive and negative
slope values at low $\beta=0.04$, while presents a positive slope for medium
and large $\beta$. The pQCD approach without saturation, again shows a
positive slope for the whole $Q^2$ and $x_{\pom}$ range. Since  the
diffractive cross section is strongly sensitive to the infrared cutoff, one
of the main differences between these models is the assumption related to the
small $Q^2$ region. In the pQCD models without saturation, an ad hoc cutoff
in the transverse momentum is inserted, as well as the energy dependence of
the non-integrated gluon distribution. In the saturation model instead,  the
saturation radius $R_0(x)$ gives the infrared cutoff (the saturation
momentum scale) and determines the energy dependence. If this scale is large
(1-2 GeV$^2$), then the resulting process is not soft and can be completely
calculated using pQCD methods. Therefore, the saturation model extends the
pQCD approach towards lower $Q^2$ values, making this region an important
source of information about the dynamics. We conclude that, the difference
between the behaviors predicted by these two models for the $x_{\pom}$
spectrum, mainly in the region of small $\beta$ and medium $Q^2$, are large,
which should allow to discriminate the dynamics in future experimental
analyzes.

For completeness we mention that a hybrid approach for the
saturation phenomena has been proposed recently mixing  the
multi-Pomeron exchange and concepts of pQCD, namely considering
the configurations in Fock space of the virtual photon
\cite{CFSK1}. In that work the $q\bar{q}$ pair is separated in
contributions of large size, described by "aligned jet"
configuration, and contributions of short size driven by
perturbative QCD as discussed in the previous sections. The gluon
emission contribution is disregarded in this case. The multiple
Pomeron exchange is taken into account considering a quasieikonal
form for the cross sections written in impact parameter
representation. Results for diffractive DIS  are  obtained since
the multiple Pomeron exchanges are related by AGK-cutting rules to
shadowing corrections to diffractive production.  Such framework
produces a good description of available experimental data from
photoproduction up to about 10 GeV$^2$.  Their conclusions are
similar to Ref. \cite{GW}, namely the saturation effects at HERA
energies are very important in the region of small $Q^2$. However,
unitarity corrections seem to appear at large $Q^2$ in
\cite{CFSK1} coming from the large size configuration and the
triple Pomeron interaction. A direct comparison between the
approaches from the Refs. \cite {GW} and \cite{CFSK1} concerning
the resultant $F_2^D$-slope is a interesting issue and it will be
postpone to a future analyzes.

\section{CONCLUSIONS}

\label{conclusao}

The study of electroproduction at small 
$x$ has lead to the improvement of our understanding of QCD dynamics in the
interface of perturbative and non-perturbative physics. However, many
important problems remain  unsolved. One of the main open questions
is if the transition between large (pQCD) and small (Regge) virtualities is
already reached or if a high density (saturation) regime is present. At this
moment, there are several approaches, based in very distinct paradigms, which
describe the current observables measured in HERA. In this paper we
propose and analyze the diffractive logarithmic slope as a potential quantity
to explicit the leading dynamics at $ep$ diffractive processes. We study in
detail the prediction for the logarithmic slope of the diffractive structure
function, using three distinct models for diffractive DIS. We choose as a pure
Regge approach the CKMT model, which describes with quite good agreement the
inclusive and diffractive structure function. The pQCD model used is a
parametrization based on the Bartels-W\"{u}sthoff approach, which describes
successfully the diffractive DIS and other observables, like jet production.
One analyzes was also performed for saturation phenomena, discussing the
results from the phenomenological approach by Golec Biernat-Wusthoff,
describing its main features and comparison with the pQCD model without
saturation. As a summary of our main results, we predict that: (a) the
analyzes of the $\beta $ spectrum in the region of small values of this
variable and the signal of the slope in the  $x_{\pom}$ spectrum at medium $%
Q^2$ values should allow to discriminate between the Regge- and pQCD-based
approaches; (b) The analyzes of the  $x_{\pom}$ spectrum of the slope at
small values of $\beta $ concerning to the $Q^2$ behavior should allow to
discriminate between the saturated and non-saturated pQCD models. 

A possible route to distinguish experimentally between the three models
discussed in this paper is the following: (i) First, analyze the signal of
the slope at 
$\beta $=0.4 and medium $Q^2$ ($\approx 10$ GeV$^2$). If this signal is
negative, then the dynamics of diffractive DIS is Regge based. On the other
hand, if the signal is positive, a pQCD approach is necessary to describe
the data; (ii) Second, to discriminate between saturated and non-saturated
models, we can analyze the  $x_{\pom}$ of the slope at small values of $%
\beta $ and medium $Q^2$. While the non-saturated pQCD model predicts a
steep growth at small values of  $x_{\pom}$, the saturated model predicts an
almost constant behavior. Of course, our results are dependent of the
assumptions used in the models considered as input in the analyzes of the
diffractive slope. However, we expect that the main point, the fact that
this quantity is a potential observable to discriminate between the hard,
semihard and soft dynamics, should not be modified by future theoretical
improvements of the diffractive DIS.

We expect that our results could motivate the experimental analyzes of the
logarithmic slope of the diffractive structure function in next years, since
the behavior of this observable may  explicit the dynamics in the small 
$x$ regime. 

\section*{ACKNOWLEDGEMENTS}

MBGD acknowledges useful discussions with A. Capella, M. Derrick and E.
Ferreiro. VPBG thanks the FAPERGS and CNPq for support. MVTM acknowledges K.
Golec-Biernat for useful enlightenments in the saturation model, as well M.
Arneodo and P. Newman for comments on H1 and ZEUS data. This
work was partially financed by CNPq and PRONEX (Programa de Apoio a N\'ucleos
de Excel\^encia), BRAZIL.

\newpage 

\end{document}